\documentclass[conference]{IEEEtran}
\usepackage[left=0.63in,right=0.625in,top=0.75in,bottom=1in]{geometry}
\usepackage[T1]{fontenc}

\usepackage{cite}
\usepackage{graphicx}
\graphicspath{{Figure/}}
\usepackage[cmex10]{amsmath}
\usepackage{amsthm}
\usepackage{amssymb}
\usepackage{algorithmic}
\usepackage{array}
\usepackage{color}
\usepackage{epstopdf}
\usepackage{amsfonts}
\usepackage{siunitx}
\usepackage{soul}
\usepackage{xurl}

\usepackage[acronym,shortcuts]{glossaries}

\usepackage{pgfplots}
\pgfplotsset{compat=1.3}
\usepackage{tikz}							% Tikz
\usetikzlibrary{shapes}
\usetikzlibrary{spy}
\usetikzlibrary{circuits}
\usetikzlibrary{arrows}	
\usepackage{subfig}

\usepackage{mathtools}

\makeglossaries
\newacronym[plural=BSs,firstplural=base stations (BSs)]{bs}{BS}{base station}
\newacronym{psd}{PSD}{power spectral density}
\newacronym[plural=PAs,firstplural=power amplifiers (PAs)]{pa}{PA}{power amplifier}
\newacronym{sdr}{SDR}{signal-to-distortion ratio}
\newacronym{aclr}{ACLR}{adjacent channel leakage ratio}
\newacronym{cpu}{CPU}{central processing unit}
\newacronym{iid}{i.i.d.}{independently and identically distributed}
\newacronym{ue}{UE}{user equipment}
\newacronym{los}{LoS}{line-of-sight}
\newacronym{nlos}{NLoS}{non-line-of-sight}
\newacronym{ula}{ULA}{uniform linear array}
\newacronym{hpbm}{HPBM}{half power beam width}
\newacronym{rts}{RTS}{ray tracing simulator}
\newacronym{ag}{AG}{array gain}
\newacronym{arp}{ARP}{Antenna Reference Point}
\newacronym{ecdf}{eCDF}{Empirical Cumulative Distribution Function}
\newacronym{evm}{EVM}{Error Vector Magnitude}
\newacronym{fft}{FFT}{fast Fourier transform}
\newacronym{ib}{IB}{in-band}
\newacronym{im}{IM}{intermodulation}
\newacronym{mimo}{MIMO}{multiple input multiple output}
\newacronym{mf}{MF}{matched filtering}
\newacronym{mr}{MR}{maximum ratio}
\newacronym{mrt}{MRT}{Maximum Ratio Transmission}
\newacronym{ofdm}{OFDM}{orthogonal frequency division multiplexing}
\newacronym{oob}{OOB}{out-of-band}
\newacronym{papr}{PAPR}{Peak-to-Average Power Ratio}
\newacronym{rx}{RX}{Receiver}
\newacronym{trp}{TRP}{Transmission Reception Point}
\newacronym{tx}{TX}{transmitter}
\newacronym{zf}{ZF}{zero forcing}
\newacronym{snr}{SNR}{signal-to-noise ratio}
\newacronym{dl}{DL}{downlink}
\newacronym{rzf}{RZF}{Regularized Zero Forcing}
\newacronym{slc}{SLC}{Spatial Leakage Suppression}
\newacronym{kpi}{KPI}{Key Performance Indicator}
\newacronym{awgn}{AWGN}{Additive White Gaussian Noise}
\newacronym{qam}{QAM}{Quadrature Amplitude Modulation}

\begin{document}

%\linespread{0.9}
\title{Spatial Distribution of Distortion due to Nonlinear Power Amplification in Distributed Massive MIMO\vspace{-0.5em}}

\author{\IEEEauthorblockN{Fran\c{c}ois Rottenberg\IEEEauthorrefmark{1}\IEEEauthorrefmark{2}\IEEEauthorrefmark{3},
		Gilles Callebaut\IEEEauthorrefmark{1},
		Liesbet Van der Perre\IEEEauthorrefmark{1}
	}
	\IEEEauthorblockA{\IEEEauthorrefmark{1}ESAT-DRAMCO, Ghent Technology Campus, KU Leuven, 9000 Ghent, Belgium
	}
	\IEEEauthorblockA{\IEEEauthorrefmark{2}ICTEAM department, Universit\'{e} catholique de Louvain, 1348 Louvain-la-Neuve, Belgium
	}
	\IEEEauthorblockA{\IEEEauthorrefmark{3}OPERA department, Universit\'{e} libre de Bruxelles, 1050 Brussels, Belgium
	}
}

\maketitle

\begin{abstract}
	Due to the nonlinearity of \glspl{pa}, the transmit signal is distorted. Previous works have studied the spatial distribution of this distortion for a central massive \gls{mimo} array. In this work, we extend the analysis for distributed massive MIMO in a \gls{los} scenario. We show that the distortion is not always uniformly distributed in space. In the single-user case, it coherently combines at the user location. In the few users case, the signals will add up at the user locations plus several others locations. As the number of users increases, it becomes close to uniformly distributed. As a comparison with a central massive MIMO system, having the same total number of antennas, the distortion in distributed massive MIMO is considerably more uniformly distributed in space. Moreover, the potential coherent combining is contained in a zone rather than in generic directions, \textit{i.e.}, in a \textit{beamspot}. Furthermore, a small-scale fading effect is observed at unintended locations due to the non-coherent combining of the signals. As a result, one can expect that going towards distributed systems allows working closer to saturation, increasing the \gls{pa} operating efficiency and/or using low-cost \glspl{pa}.

\end{abstract}


\section{Introduction}\label{section:Introduction}

Distributed  and  cell-free  massive  MIMO  are  prime  candidates  for  next  generation  wireless  systems  thanks  to  their potential  to  offer  consistent  good  service  and  increased  and scalable  capacity~\cite{NgoCellFree}.  A  highly  energy  efficient  system  can be  realized  provided  the  system  can  operate  the \glspl{pa} close to saturation. In this region however, they exhibit nonlinear behaviour generating  distortion,  both \gls{ib} and  \gls{oob}.  

The occurrence and impact of distortions due to nonlinear power amplification has been studied in literature for the case of central massive MIMO with one large central array at the basestation. When the distortions generated by the many \glspl{pa} are assumed to be uncorrelated, it was shown that they do not combine coherently in space. Hence, thanks to the large number of transmit antennas, great performance and energy efficient operation can be achieved~\cite{bjornson2014}. However, in general the distortion terms are correlated over the transmit antennas and they may significantly impact both the \gls{ib} signal quality and the \gls{oob} radiation~\cite{Larsvdp2018} under specific conditions. The spatial  distribution  of  the  distortion  due  to  nonlinear power  amplification  has  been rigorously  studied  for the case of massive MIMO central large array in~\cite{Mollen2018TWC}. The authors prove that in particular the transmission scenarios with one or few \glspl{ue} that are in \gls{los} to the basestation need to be considered. Indeed not only the wanted information signal for the intended \glspl{ue} yet also the distortions may in this case get significant antenna array gain. 

\begin{figure}[!t]  
	\centering
	\includegraphics[width=0.9\linewidth]{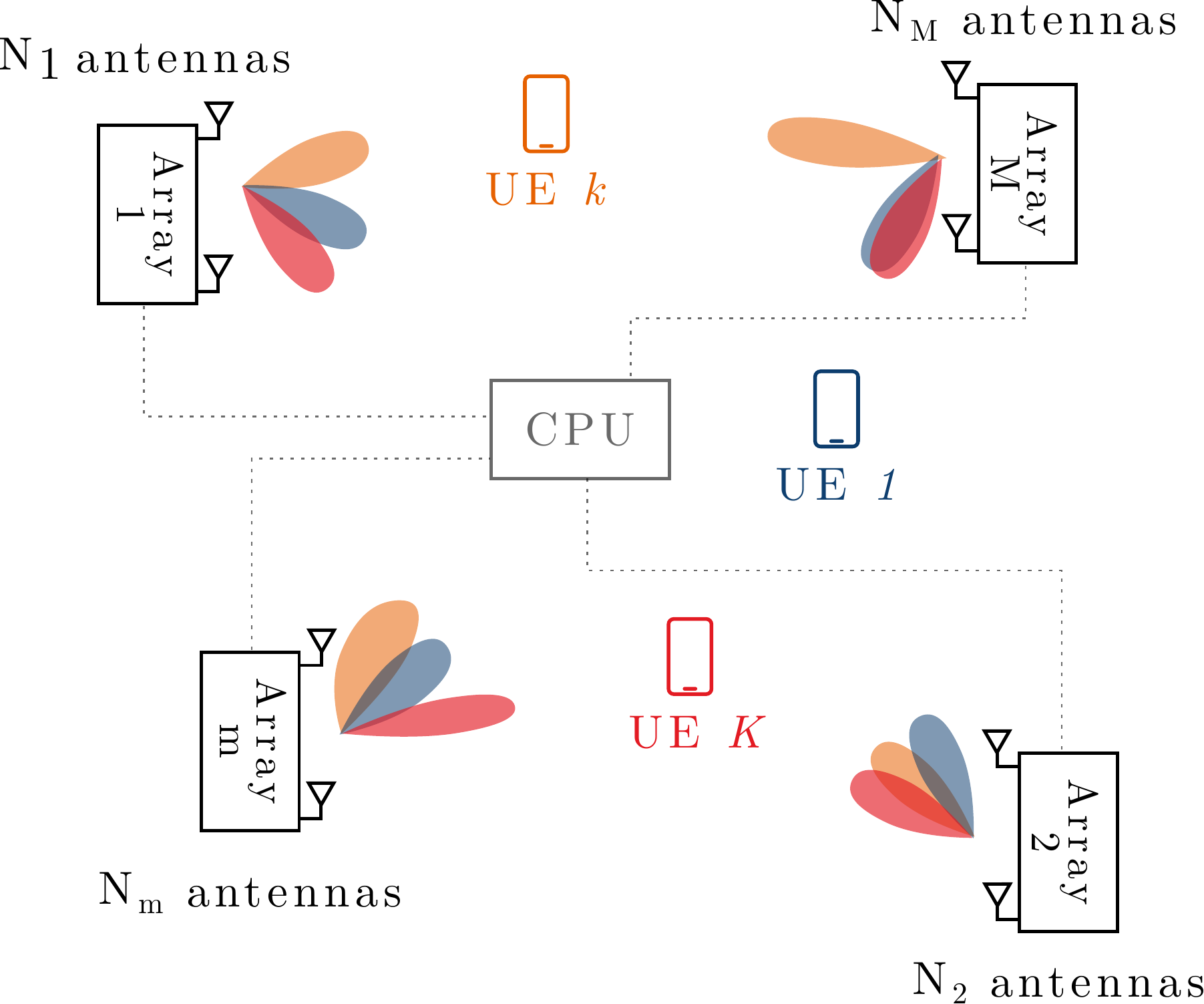} 
	%	\vspace{-1em}
	\caption{\small Distributed MIMO System with \gls{cpu}, \(K\) single-antenna users, \(M\) arrays, each with \(N_m\) antennas.}
	\label{fig:system_model}
	\vspace{-1.5em}
\end{figure}

In this paper we study the spatial distribution of distortion due to nonlinear power amplification for distributed array deployments, as depicted in Fig.~\ref{fig:system_model}, and exemplary for cell-free massive MIMO.  For simplicity and to gain insight we focus on relatively simple scenarios: antenna arrays with omnidirectional antennas and no coupling between them are considered, transmitting to a single-user or multiple users in \gls{los}. We further show that the \gls{los} conditions present a worst case in terms of detrimental combining of the distortion terms. Analytical expressions are derived for the case of a \gls{mf} beamformer and a general single-carrier formulation. We further comment on the expectations for more complex transmission scenarios.
%The files to generate the figures are open-source\footnote{\url{github.com/dramco/Spatial-Distribution-of-Distortion-due-to-Nonlinear-Power-Amplification-in-Distributed-Massive-MIMO}}.
%\linespread{0.75}	\francois{I would put that with the journal paper}

\textbf{Notations}: %\subsection*{Notations}
Superscript $^*$ stands for conjugate, $\jmath$ is the imaginary unit and $\mathbb{E}(.)$ is the expectation. $\circledast$ denotes the convolution. $\delta_n$ and $\delta(t)$ are the Kronecker and Dirac deltas respectively.%$\mat{I}_n$ denotes the identity matrix of order $n$. %$\otimes$ stands for the Kronecker product and $\delta_n$ and 
%$\delta(t)$ is the Dirac delta. %The $\diag(.)$ operator applied to a vector returns a diagonal matrix whose $k-$th diagonal entry is equal to the $k-$th entry of the argument vector. %The $\diag(.)$ operator applied to a square matrix returns the same matrix with off-diagonal elements set to zero. Operator $\ceil{.}$ is the ceiling function. 

\section{System Model}
\label{section:transmission_model}

% \begin{figure}[!t]
% 	\centering
% 	\resizebox{0.8\textwidth}{!}{%
% 		{\includegraphics[clip, trim=0cm 7cm 6cm 0cm, scale=1]{system_model.pdf}} %gauche bas droite haut
% 	}
% 	%	\vspace{-1em}
% 	\caption{Distributed MIMO System with a central processing unit (CPU), $K$ single-antenna users, $M$ arrays, each equipped with $N_m$ antennas.}
% 	\label{fig:system_model}
% 		\vspace{-1em}
% \end{figure}

\subsection{Signal Model}

As shown in Fig.~\ref{fig:system_model}, we consider a distributed massive MIMO systems with a total of $M$ antenna arrays, perfectly synchronized with one another and with perfect channel state information. We denote the number of antennas of array $m$ as $N_m$. A total of $K$ single-antenna \glspl{ue} are being served simultaneously using the same time and frequency resources. A conventional massive MIMO system with one large central array is obtained in the particular case $M=1$.

\begin{figure}[!t]  
	\centering
	
	\resizebox{0.6\textwidth}{!}{%
		{\includegraphics[clip, trim=0cm 10cm 14cm 0cm, scale=1]{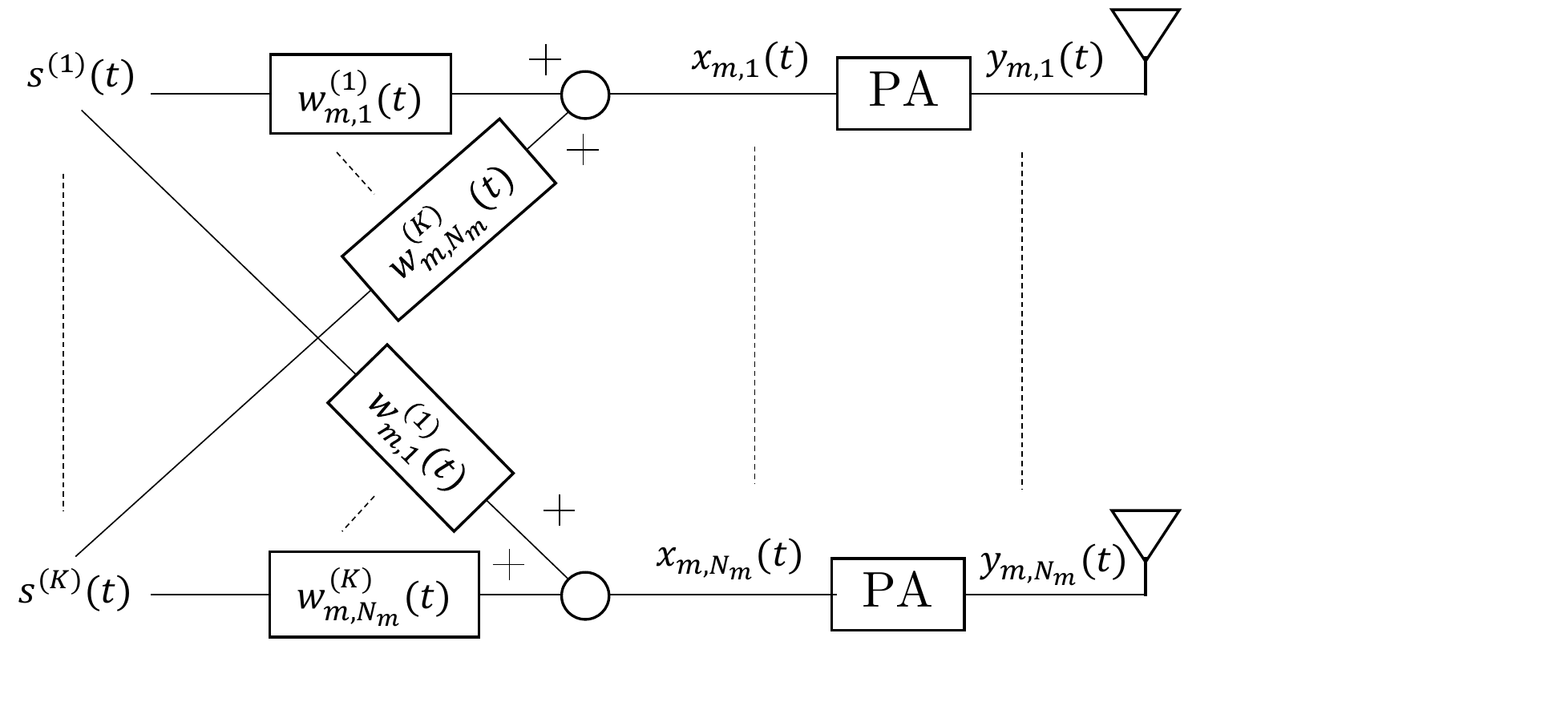}} %gauche bas droite haut
	}
	\vspace{-1.5em}
	\caption{%\gilles{redraw in inkscape} 
		Transmitter structure of array $m$ with \(N_m\) antennas. %Signals are in complex baseband representations.
	}
	\label{fig:BS_m}
	\vspace{-1.5em}
\end{figure}

The complex baseband representation of the signal intended for \gls{ue} $k$ is denoted by $s^{(k)}(t)$, for $k=1,...,K$. It is assumed to be wide sense stationary and uncorrelated between different users. They have the same \gls{psd} $S(\omega)$, with a power scaling $p^{(k)}$. We define the inverse Fourier transform of $S(\omega)$ as $R(\tau)$. The cross-correlation and cross-\gls{psd} of the transmit signals are then given by
\begin{align}
{R}_{s^{(k)}s^{(k')}}(\tau)&=\mathbb{E}\left(s^{(k)}(t+\tau)(s^{(k')}(t))^*\right)= \delta_{k-k'} p^{(k)}R(\tau)\nonumber\\ 
S_{s^{(k)}s^{(k')}}(\omega)&= \delta_{k-k'} p^{(k)} S(\omega),\label{eq:R_sksk'}
\end{align}
where we used the fact that user signals are uncorrelated. The transmitter structure of array $m$ is shown in Fig.~\ref{fig:BS_m}. The signal $s^{(k)}(t)$ is precoded at transmit antenna $n$ of array $m$  using a filter $w_{m,n}^{(k)}(t)$. The complex baseband representation of the signal before the \gls{pa} of the corresponding antenna is denoted by $x_{m,n}(t)$ and is given by
\begin{align}
x_{m,n}(t)=\sum_{k=1}^{K} s^{(k)}(t) \circledast w_{m,n}^{(k)}(t). \label{eq:x_mn}
\end{align}  
%Its passband representation is denoted by $x_{m,n}^{\mathrm{pb}}(t)$ and is given by $x_{m,n}^{\mathrm{pb}}(t)=2\Re(x_{m,n}(t)e^{\jmath \omega_c t})=x_{m,n}(t)e^{\jmath \omega_c t}+x_{m,n}^*(t)e^{-\jmath \omega_c t}$, where $\omega_c=2\pi f_c$, with $f_c$ being the carrier frequency. The signal $x_{m,n}^{\mathrm{pb}}(t)$ is fed to a nonlinear \gls{pa}. Finally, the output of the \gls{pa}, defined as $y^{\mathrm{pb}}_{m,n}(t)$, is fed to antenna $n$ of array $m$. We also define the complex baseband representation of $y^{\mathrm{pb}}_{m,n}(t)$ as $y_{m,n}(t)$.

\subsection{Nonlinear \gls{pa} Model}

Keeping only terms that appear around the carrier frequency, the baseband representation of the signal after the \gls{pa} can generally be written as \cite{Horlin2008,Mollen2018TWC}
\begin{align}
y_{m,n}(t)=\sum_{l=1,3,5,...}^{} b_l x_{m,n}(t)  |x_{m,n}(t)|^{l-1}. \label{eq:y_mn_gen}
\end{align}
The first order term ($l=1$) will be referred to as the useful signal part, while the sum of the higher order terms will be referred to as distortion. Note that distortion is present both in-band and out-of-band.
We define the cross-correlations of the signals $x_{m,n}(t)$ and $y_{m,n}(t)$ (before and after the \gls{pa}) as ${R}_{{x}_{m,n}{x}_{m',n'}}(\tau)$ and ${R}_{{y}_{m,n}{y}_{m',n'}}(\tau)$ respectively.
% \begin{align*}
% 	{R}_{{x}_{m,n}{x}_{m',n'}}(\tau)&=\mathbb{E}\left({x}_{m,n}(t+\tau){x}_{m',n'}^*(t)\right)\\
% 	{R}_{{y}_{m,n}{y}_{m',n'}}(\tau)&=\mathbb{E}\left({y}_{m,n}(t+\tau){y}_{m',n'}^*(t)\right).
% \end{align*}
As in \cite{Mollen2018TWC}, we assume that the complex baseband signal $x_{m,n}(t)$ can be modeled as a zero mean circularly symmetric Gaussian random variable. This is particularly true if the signal is beamformed and OFDM modulated.\footnote{We here consider a general single carrier signal but the study can be straightforwardly extended to a multicarrier modulated signal.} %\liesbet{here we are not consistent with what was said in the introduction, or at least confusing, we could clarify main carrier frequency vs sub-carriers}\francois{I see what you mean but you could actually see an OFDM system as a particular case of our formulation. In other words, we consider a general passband transmission of a signal $s(t)$ around a carrier frequency $f_c$. We do not particularize to a multicarrier system but it could be done (with some modifications...). Let me know what you think of the footnote I added, I try to "milquetiseren" a bit}.
Then, we can use the result of \cite{Mollen2018TWC,phdthesisPaperE} to relate the cross-antenna correlation of the input of the \gls{pa} to the one at the output of the \gls{pa}
\begin{align}
&{R}_{{y}_{m,n}{y}_{m',n'}}(\tau)\label{eq:R_yy}\\
&= \sum_{l=1,3,5,...}^{} c_l {R}_{{x}_{m,n}{x}_{m',n'}}(\tau) |{R}_{{x}_{m,n}{x}_{m',n'}}(\tau)|^{l-1},\nonumber
\end{align}
where $n$ and $n'$ are antenna indices of array $m$ and $m'$ respectively. Coefficients $c_l$ can be related to $b_l$ and the input power of each PA~\cite{Mollen2018TWC}.\footnote{Coefficients $c_l$ here do not depend the PA index $m,n$ since all are working in the same power regime, as will be shown in Section~\ref{section:spatial_distribution}.} %Equation~(\ref{eq:R_yy}) will prove useful to study the spatial distribution of the distortion in following sections.

\subsection{Channel Model}

The channel from each array to a given location is assumed in pure \gls{los} with no other multipath components, and in the far field from each array. %\footnote{This presents a worst case scenario \gilles{for what, worst-case/wrt which metric?} in view of operating PAs in an energy efficient mode close to saturation. The situations with a strong dominant \gls{los} will yield similar results\francois{I agree with this comment, good that you inserted it, maybe in the main body and not in footnote? we could also discuss it in the latest section regarding NLOS conditions and MPC}\TODO{add comment in intro}}. 
This location could be either a \gls{ue} or an observer location. Considering omnidirectional antennas and \glspl{ula}, the baseband representation of the channel impulse response between the given location and antenna $n$ of array $m$ is
\begin{align*}
{h}_{m,n}(t) &= \beta_m \delta(t-\tau_{m}) e^{-\jmath \phi_m n}.
\end{align*}
The delay $\tau_{m}$ accounts for the propagation delay between array $m$ and the given location. The complex coefficient $\beta_m=|\beta_m|e^{-\jmath \psi_m}$, models the path loss between array $m$ and the given location together with a phase shift $\psi_m=2\pi f_c \tau_m$. Finally, the complex exponential $e^{-\jmath \phi_m n}$ accounts for the phase shift of antenna $m$ with respect to reference antenna $n=0$, with $\phi_m=\frac{2\pi}{\lambda_c} d\cos(\theta_m)$, where $\lambda_c$ the carrier wavelength, $d$ the inter-antenna spacing and $\theta_m$ the angular direction of the given location with respect to array $m$. Moreover, we define the channel between the \gls{ue} $k$ and antenna $n$ of array $m$ as
\begin{align*}
{h}_{m,n}^{(k)}(t) &= \beta_m^{(k)} \delta(t-\tau_{m}^{(k)}) e^{-\jmath \phi_m^{(k)} n}.
\end{align*}
The received signal at the considered location is given by
\begin{align}
r(t) &= \sum_{m=1}^{M}\sum_{n=0}^{N_m-1} {y}_{m,n}(t) \circledast {h}_{m,n}(t)\nonumber\\
&= \sum_{m=1}^{M}\beta_m\sum_{n=0}^{N_m-1} {y}_{m,n}(t-\tau_{m}) e^{-\jmath \phi_m n}.\label{eq:r_gen}
\end{align}

\section{Spatial Distribution of Signal and Distortion}
\label{section:spatial_distribution}

The spatial distribution of the nonlinear distortion depends on the type of beamformer being used. Given its practical convenience in the distributed setting, we here consider a matched filter
\begin{align}
w_{m,n}^{(k)}(t) = \frac{(h_{m,n}^{(k)}(-t))^*}{\|h_{m,n}^{(k)}(-t)\|}=  \delta(t+\tau_{m}^{(k)}) e^{\jmath \phi_m^{(k)} n+\jmath\psi_m^{(k)}}. \label{eq:MF_beamformer}
\end{align}
This precoding will ensure that the signals transmitted by each antenna of each array will add up coherently at \gls{ue} $k$. We have considered a uniform per-array power allocation, which avoids the need for coordination. The power is also uniform over the antennas, given the constant LoS channel gain $|\beta_m^{(k)}|$ across the antennas. Inserting (\ref{eq:MF_beamformer}) into (\ref{eq:x_mn}), the input of the \gls{pa} becomes
\begin{align}
x_{m,n}(t)&=\sum_{k=1}^{K} s^{(k)}(t+\tau_{m}^{(k)}) e^{\jmath \phi_m^{(k)} n+\jmath\psi_m^{(k)}}. \label{eq:x_mf}
\end{align}  
Using (\ref{eq:R_sksk'}), its cross-correlation is given by
\begin{align}
&{R}_{{x}_{m,n}{x}_{m',n'}}(\tau)\label{eq:R_x_x_sp}\\
&= \sum_{k=1}^{K} p^{(k)}e^{\jmath (\phi_m^{(k)}n-\phi_{m'}^{(k)}n'+\psi_m^{(k)}-\psi_{m'}^{(k)})} R(\tau+\tau_{m}^{(k)}-\tau_{m'}^{(k)}).\nonumber
\end{align}
One can note that all \glspl{pa} are working in the same saturation regime since the power at each \gls{pa} is the same, \textit{i.e.}, $\sigma_{{x}_{m,n}}^2={R}_{{x}_{m,n}{x}_{m,n}}(0)=\sum_{k=1}^{K} p^{(k)}R(0),\ \forall m,n$. The cross-correlation of the amplified signals, ${R}_{{y}_{m,n}{y}_{m',n'}}(\tau)$, can be computed inserting this last result into (\ref{eq:R_yy}). Using (\ref{eq:r_gen}), the autocorrelation of the received signal $r(t)$ is then obtained as
\begin{align}
R_{rr}(\tau)%=\mathbb{E}\left(r(t+\tau)(r(t))^*\right)\nonumber\\
&= \sum_{m=1}^{M}\sum_{n=0}^{N_m-1}\sum_{m'=1}^{M}\sum_{n'=0}^{N_{m'}-1} \beta_m\beta_{m'}^*  e^{-\jmath (\phi_m n-\phi_{m'} n')}\nonumber\\
&\quad \quad {R}_{{y}_{m,n}{y}_{m',n'}}(\tau-(\tau_{m}-\tau_{m'})). \label{eq:R_rr}%\\
%&= \sum_{m,n,m',n'} \beta_m\beta_{m'}^*  e^{-\jmath (\phi_m n-\phi_{m'} n')}{R}_{{y}_{m,n}{y}_{m',n'}}(\tau-(\tau_{m}-\tau_{m'})). \label{eq:R_rr}
\end{align}
Finally, the \gls{psd} of the signal received at any given location can be obtained through a Fourier transform as
\begin{align}
&{S}_{rr}(\omega)%&=\int_{-\infty}^{+\infty} R_{rr}(\tau) e^{-\jmath \omega \tau} d\tau \\
%&=\sum_{m=1}^{M}\sum_{n=0}^{N_m-1}\sum_{m'=1}^{M}\sum_{n'=0}^{N_{m'}-1} \sqrt{\beta_m}\sqrt{\beta_{m'}}  e^{-\jmath (\phi_m n-\phi_{m'} n')}\nonumber\\
%&\int_{-\infty}^{+\infty}{R}_{{y}_{m,n}{y}_{m',n'}}(\tau-(\tau_{m}-\tau_{m'}))e^{-\jmath \omega \tau} d\tau\nonumber\\
=\sum_{m=1}^{M}\sum_{n=0}^{N_m-1}\sum_{m'=1}^{M}\sum_{n'=0}^{N_{m'}-1} \beta_m\beta_{m'}^*  e^{-\jmath (\phi_m n-\phi_{m'} n')}\nonumber\\
&e^{-\jmath \omega (\tau_{m}-\tau_{m'})}\int_{-\infty}^{+\infty}{R}_{{y}_{m,n}{y}_{m',n'}}(\tau')e^{-\jmath \omega \tau'} d\tau'\label{eq:S_rr}.
%&=\sum_{m=1}^{M}\sqrt{\beta_m}e^{-\jmath \omega \tau_{m}}\sum_{n=0}^{N_m-1}e^{-\jmath \phi_m n}\sum_{m'=1}^{M}\sqrt{\beta_{m'}}e^{\jmath \omega \tau_{m'}}\sum_{n'=0}^{N_{m'}-1}   e^{\jmath \phi_{m'} n'}\nonumber\\
%&\int_{-\infty}^{+\infty}{R}_{{y}_{m,n}{y}_{m',n'}}(\tau)e^{-\jmath \omega \tau} d\tau\nonumber.
\end{align}
%where we used (\ref{eq:R_rr}) and a change of variable $\tau'=\tau-(\tau_{m}-\tau_{m'})$. 
The computation of ${S}_{rr}(\omega)$ can be easily computed numerically but is tedious to write down analytically. In the following sections, we particularize the problem to the single-user case to get more insight. Then, we focus on the multi-user case and we study the beam directions of the signal and the third-order term of the distortion.

\subsection{Single-User Case distributed array deployment}

In this case, $K=1$ and the signal before the \gls{pa} (\ref{eq:x_mf}) becomes
\begin{align*}
x_{m,n}(t)&=s^{(1)}(t+\tau_{m}^{(1)}) e^{\jmath \phi_m^{(1)} n+\jmath\psi_m^{(1)}}.
\end{align*}
The signal after the \gls{pa} (\ref{eq:y_mn_gen}) becomes
\begin{align}
&y_{m,n}(t)\label{eq:y_mn_SU}\\
&=e^{\jmath \phi_m^{(1)} n+\jmath\psi_m^{(1)}} \sum_{l=1,3,5,...}^{} b_l s^{(1)}(t+\tau_{m}^{(1)}) \left|s^{(1)}(t+\tau_{m}^{(1)})\right|^{l-1}, \nonumber
\end{align}
which shows that both the signal and the distortion are beamformed in the same direction from each individual array, \textit{i.e.}, the \gls{ue} direction $\phi_m^{(1)}$. To be rigorous, the angle $\phi_m^{(1)}$ is a phase shift that can be mapped to a physical angular direction $\theta$ at each array as $\phi=\frac{2\pi}{\lambda_c} d\cos(\theta) \Longleftrightarrow \theta = \arccos (\phi\lambda_c/(2\pi d))$. The received signal (\ref{eq:r_gen}) then becomes
\begin{align*}
&r(t) %&= \sum_{m=1}^{M}\sqrt{\beta_m}\sum_{n=0}^{N_m-1} {y}_{m,n}(t-\tau_{m}) e^{-\jmath \phi_m n}\\
=\sum_{m=1}^{M}|\beta_m|e^{\jmath (\psi_m^{(1)}- \psi_m)}D_{N_m}(\phi_m^{(1)}- \phi_m)\\%\sum_{n=0}^{N_m-1} e^{\jmath (\phi_m^{(1)}- \phi_m) n} \\
&\sum_{l=1,3,5,...}^{} b_l s^{(1)}(t-(\tau_{m}-\tau_{m}^{(1)})) \left|s^{(1)}(t-(\tau_{m}-\tau_{m}^{(1)}))\right|^{l-1},
\end{align*}
where $D_{N}(\phi)=\sum_{n=0}^{N-1} e^{\jmath \phi n}$ has the shape of a Dirichlet kernel. Note that at the \gls{ue} location, \textit{i.e.}, $\tau_{m}=\tau_{m}^{(1)}$, $\psi_{m}=\psi_{m}^{(1)}$ and $\phi_m=\phi_m^{(1)}$, the angle and delay mismatches cancel. As a result, both the signal and the distortion add up coherently in an area around the \gls{ue}. We call this area where the beamformed distortions from the different arrays coherently combine a 'beamspot'.  % \gilles{I would clarify why we use the term beamspot rather than beam. This to further highlight the fact that it is a distributed case.} 
Let us now look at the \gls{psd} of the received signal. Equation (\ref{eq:R_x_x_sp}) can be particularized to the single-user case ($K=1$). Inserting this result into (\ref{eq:R_yy}) allows to evaluate ${R}_{{y}_{m,n}{y}_{m',n'}}(\tau)$. Inserting this result into (\ref{eq:S_rr}) finally gives, after a few mathematical manipulations,
\begin{align}
&{S}_{rr}(\omega)=\left(S_{\mathrm{sig}}(\omega) + S_{\mathrm{dis}}(\omega)\right)\label{eq:S_rr_SU}\\
&\left|\sum_{m=1}^{M}|\beta_m|e^{\jmath (\psi_m^{(1)}- \psi_m)-\jmath \omega (\tau_{m}-\tau_{m}^{(1)})} D_{N_m}(\phi_m^{(1)}- \phi_m)\right|^2, \nonumber
\end{align}
where we defined the \gls{psd} of the useful signal and distortion terms respectively as $S_{\mathrm{sig}}(\omega)=c_1 p^{(1)} S(\omega),\ S_{\mathrm{dis}}(\omega)=\sum_{l=3,5,...}^{}  c_l (p^{(1)})^l S^{(l)}(\omega)$, with
% \begin{align*}
% 	S_{\mathrm{sig}}(\omega)&=c_1 p^{(1)} S(\omega),\ S_{\mathrm{dis}}(\omega)=\sum_{l=3,5,...}^{}  c_l (p^{(1)})^l S^{(l)}(\omega)\\%,
%\end{align*}
%with the following definition of the \gls{psd} of the $l$-th order term of the distortion
\begin{align*}
S^{(l)}(\omega)&= \int_{-\infty}^{+\infty} R(\tau) |R(\tau)|^{l-1} e^{-\jmath \omega \tau} d\tau.
\end{align*}
%\begin{figure}[t]
%	\centering 
%	\resizebox{0.4\textwidth}{!}{\large \input{PSD}}
%	\caption{\small \Glspl{psd} of the signal $s(t)$ and the three first distortion orders, for a raised cosine filter with roll-off factor $0.2$, of symbol period $T$. All \glspl{psd} are normalized to have the same power. Note that in reality higher order terms will exhibit significantly lower power levels.}%
%	\label{fig:PSD}
%	%	\vspace{-2.5em}
%\end{figure}
%An example of these \glspl{psd} is shown in Fig.~\ref{fig:PSD}. 
Again, equation~(\ref{eq:S_rr_SU}) clearly shows that both the \gls{psd} of the useful signal $S_{\mathrm{sig}}(\omega)$ and the distortion $S_{\mathrm{dis}}(\omega)$ are affected by the same spatial gain. Hence, they will have exactly the same spatial distribution. Moreover, only at the location of the \gls{ue} will the signals coherently combine, elsewhere the signals originating from the different antennas will in general not add up coherently. At the \gls{ue} location, (\ref{eq:S_rr_SU}) becomes
\begin{align}
{S}_{rr}(\omega)%&=\int_{-\infty}^{+\infty} R_{rr}(\tau) e^{-\jmath \omega \tau} d\tau.\\
%&=\sum_{m=1}^{M}\sqrt{\beta_m}\sum_{n=0}^{N_m-1}e^{-\jmath (\phi_m-\phi_m^{(1)}) n}\sum_{m'=1}^{M} \sqrt{\beta_{m'}}\sum_{n'=0}^{N_{m'}-1}   e^{\jmath (\phi_{m'} -\phi_m^{(1)}) n'}\nonumber\\
%&e^{-\jmath \omega \left((\tau_{m}-\tau_{m}^{(1)})-(\tau_{m'}-\tau_{m'}^{(1)})\right)}\sum_{l=1,3,5,...}^{}  p^l S^{(l)}(\omega) 
=&\left(S_{\mathrm{sig}}(\omega) + S_{\mathrm{dis}}(\omega)\right)(\sum_{m=1}^{M}|\beta_m^{(1)}| N_m )^2\label{eq:S_rr_SU_user_location}
,
\end{align}
which shows that both signal and distortion add up coherently with a maximal array gain.

\begin{figure*}[t!]
	\centering 
	\subfloat[$K=2$, $N=32$.
	]{
		\resizebox{0.23\linewidth}{!}{%
			{\includegraphics[clip, trim=3.4cm 2.6cm 3.1cm 3cm, scale=1]{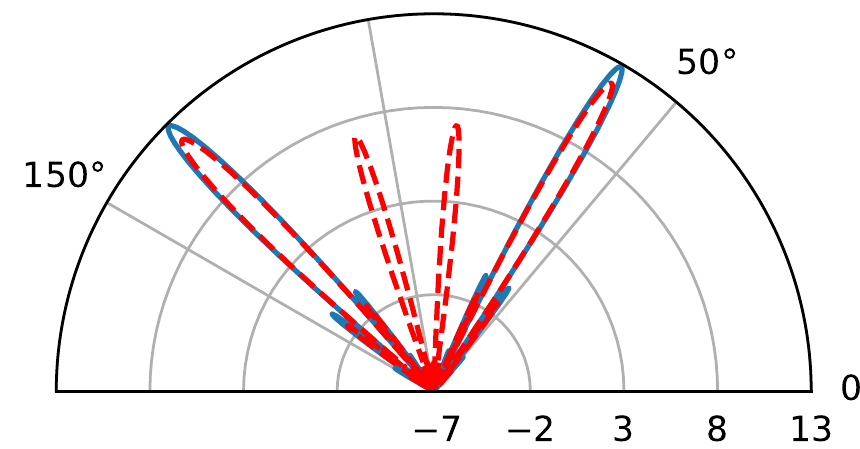}} %gauche bas droite haut
		}
	} %\hspace{-1em}
	\subfloat[$K=2$, $N=8$.]{
		\resizebox{0.23\linewidth}{!}{%
			{\includegraphics[clip, trim=3.4cm 2.6cm 3.1cm 3cm, scale=1]{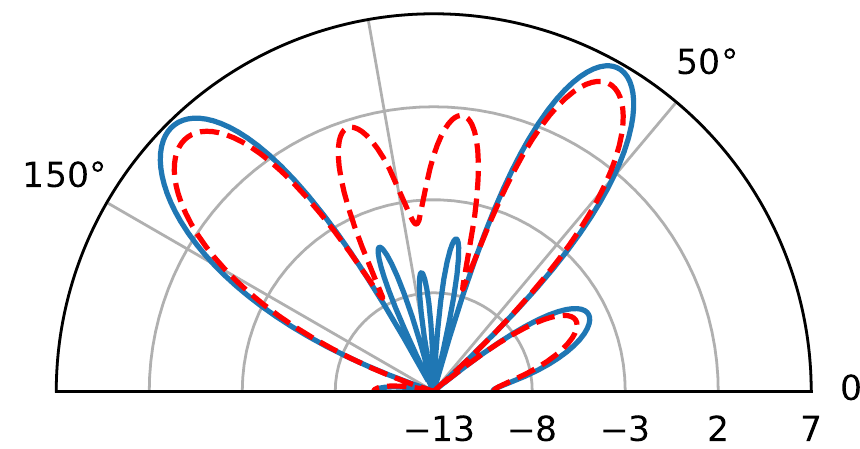}} %gauche bas droite haut
		}
	}
	\subfloat[$K=3$, $N=32$.]{
		\resizebox{0.23\linewidth}{!}{%
			{\includegraphics[clip, trim=3.4cm 2.6cm 3.15cm 3cm, scale=1]{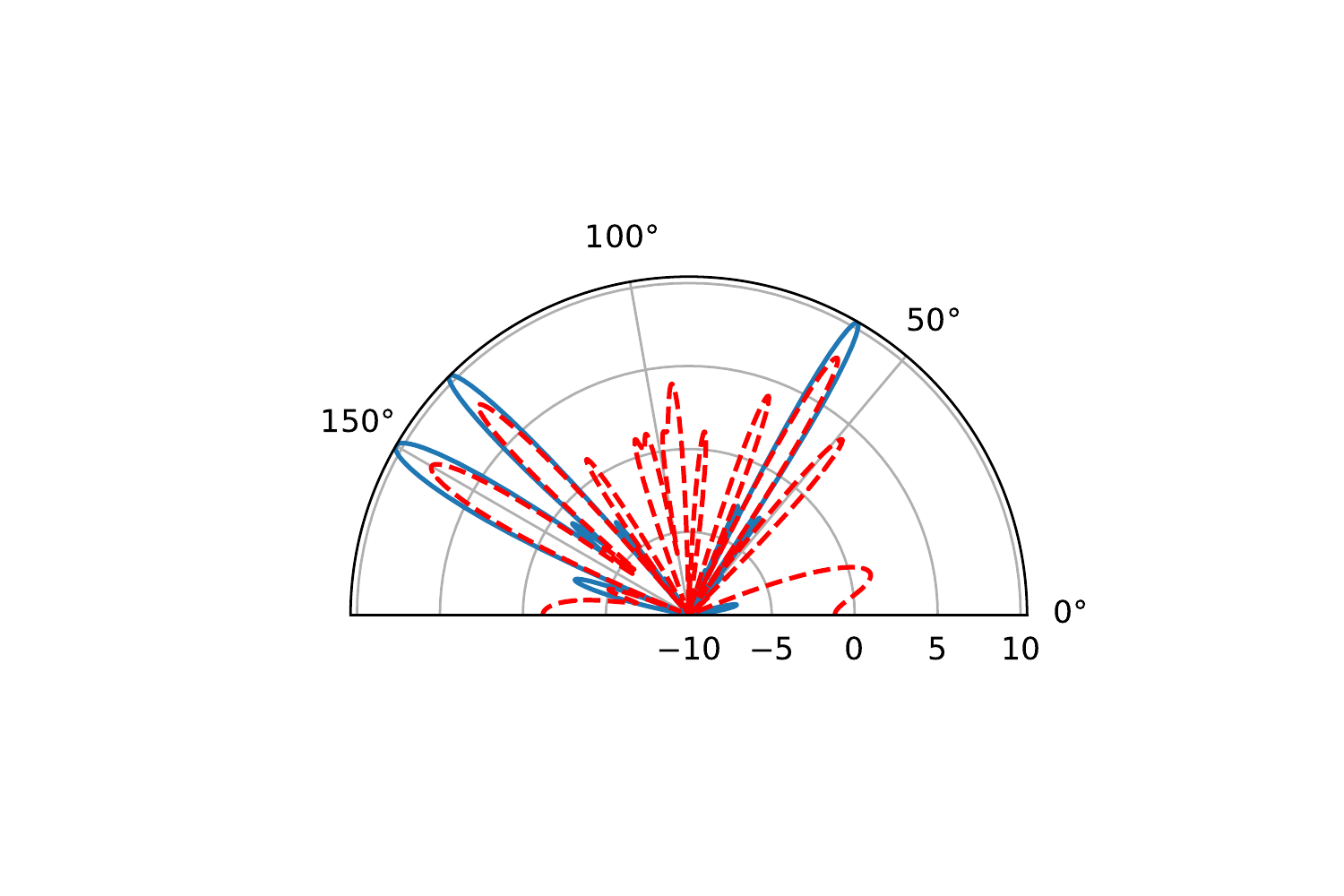}} %gauche bas droite haut
		}
	}
	\subfloat[$K=3$, $N=8$.]{
		\resizebox{0.23\linewidth}{!}{%
			{\includegraphics[clip, trim=3.4cm 2.6cm 3.15cm 3cm, scale=1]{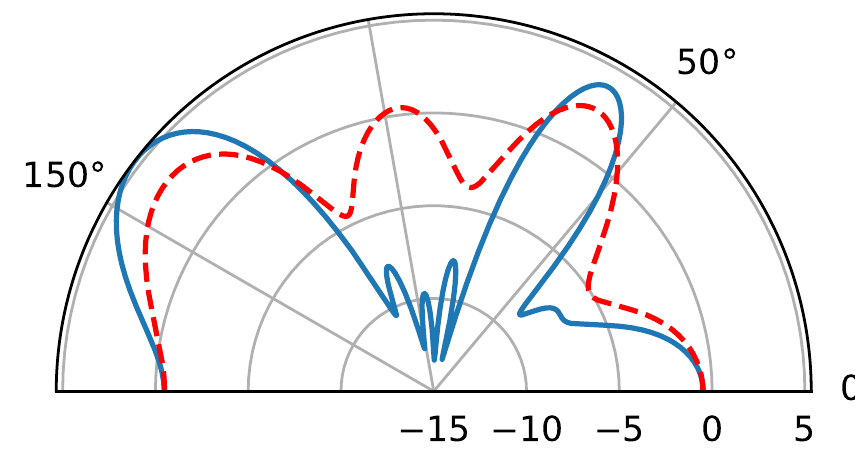}} %gauche bas droite haut
		}
	}
	\caption{\small Array directivity [dB] of the useful signal (blue) and third-order distortion (dashed red). Equal power is sent to each user and $d=\lambda_c/2$. User angles $\theta$ are: 135\textdegree , 60\textdegree\ and 150\textdegree.  %Note that the maximum directivity is different in the four shown cases. 
		As the number of users ($K$) increases, more intermodulation beams appear, the distortion becomes more uniformly distributed and the ratio of the array gain for the useful signal versus the distortion increases. Reducing the number of antennas ($N$) at the array induces a large beam width and thus a more uniform distribution.
	}
	\label{fig:radiation_patterns} 
	\vspace{-1em}
\end{figure*}

\subsection{Comparison of central vs distributed for the single-user case}

The performance of a central massive MIMO system can be retrieved by setting $M=1$. In that case, we can drop the index $m$ and (\ref{eq:S_rr_SU}) simplifies to
\begin{align}
{S}_{rr}(\omega)%&=\int_{-\infty}^{+\infty} R_{rr}(\tau) e^{-\jmath \omega \tau} d\tau.\\
%&=\sum_{m=1}^{M}\sqrt{\beta_m}\sum_{n=0}^{N_m-1}e^{-\jmath (\phi_m-\phi_m^{(1)}) n}\sum_{m'=1}^{M} \sqrt{\beta_{m'}}\sum_{n'=0}^{N_{m'}-1}   e^{\jmath (\phi_{m'} -\phi_m^{(1)}) n'}\nonumber\\
%&e^{-\jmath \omega \left((\tau_{m}-\tau_{m}^{(1)})-(\tau_{m'}-\tau_{m'}^{(1)})\right)}\sum_{l=1,3,5,...}^{}  p^l S^{(l)}(\omega) \\
=&\left(S_{\mathrm{sig}}(\omega) + S_{\mathrm{dis}}(\omega)\right)|\beta|^2 |D_{N}(\phi^{(1)}- \phi)|^2\label{eq:S_rr_SU_massive_MIMO}.
\end{align}
Comparing (\ref{eq:S_rr_SU}) and (\ref{eq:S_rr_SU_massive_MIMO}), we can observe that in central massive MIMO, the spatial gain is determined by the angular direction of the given location $\phi$ with respect to the array, together with the path loss $|\beta|^2$, which is mainly distance dependent. In distributed massive MIMO, the spatial gain depends on more parameters, which makes it focused on a 'beamspot' space: i) The path loss coefficient $\beta_m$ with respect to each array. ii) the directions $\phi_m$ with respect to each array. iii) The propagation delay $\tau_m$ with respect to each array. iv) The phase shift $\psi_m=2\pi f_c \tau_m$. The fact that the signal coming from each array will arrive with a different timing offset will induce a combining that is in general non-coherent at unintended locations. This effect can be seen as small scale fading since it varies with displacement on the order of the wavelength and is highly frequency dependent. This fading will become Rayleigh distributed as the number of arrays grows large and their contributions are similar. The particularity here is that the Rayleigh fading is not due to $M$ reflective components of a single transmit signal, since the channel is in pure \gls{los}, yet originates from the non-coherent transmissions from $M$ arrays at the unintended locations. If the location is much closer to one specific array $m$, its contribution can be much stronger than the contributions of other arrays. In that case, the fading will look more like Rician fading.

\begin{figure*}[t!]
	\centering 
	\resizebox{1\linewidth}{!}{%
		{\includegraphics{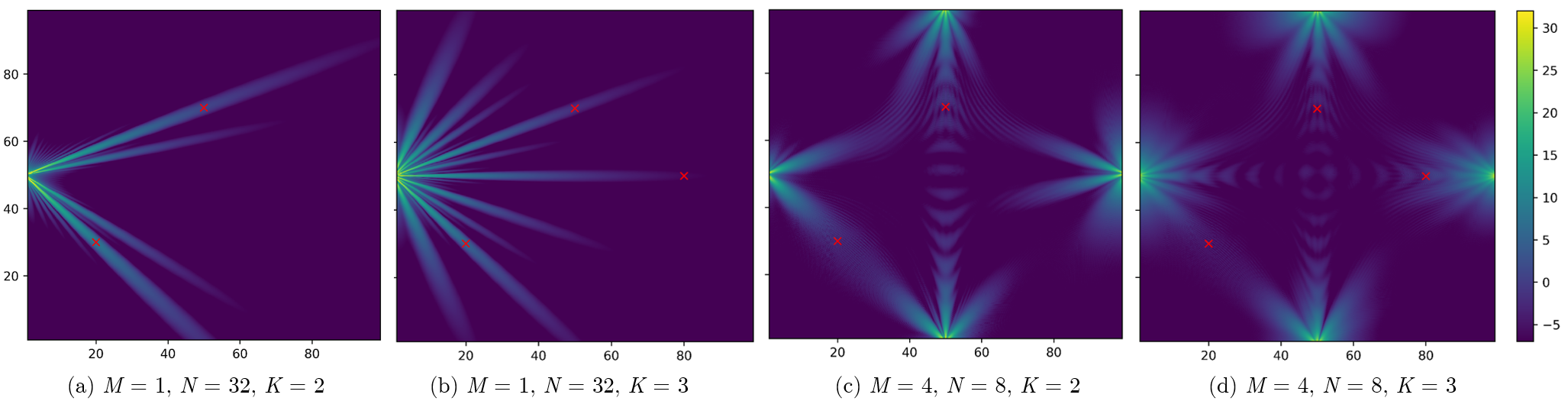}} %gauche bas droite haut
	}
	%	\subfloat[$M=1$, $N=32$, $K=2$.
	%	]{
	%		\resizebox{0.35\linewidth}{!}{%
	%		{\includegraphics[clip, trim=1cm 0.5cm 2cm 1cm, scale=1]{distribution_dist_M_1_N_32_K_2.png}} %gauche bas droite haut
	%	}
	%	} %\hspace{-1em}
	%	\subfloat[$M=4$, $N=8$, $K=2$.]{
	%		\resizebox{0.35\linewidth}{!}{%
	%		{\includegraphics[clip, trim=1cm 0.5cm 1cm 1cm, scale=1]{distribution_dist_M_4_N_8_K_2.png}} %gauche bas droite haut
	%	}
	%	}
	%	
	%	
	%	\subfloat[$M=1$, $N=32$, $K=3$.]{
	%		\resizebox{0.35\linewidth}{!}{%
	%		{\includegraphics[clip, trim=1cm 0.5cm 1cm 1cm, scale=1]{distribution_dist_M_1_N_32_K_3.png}} %gauche bas droite haut
	%	}
	%	} %\hspace{-1em}
	%	\subfloat[$M=4$, $N=8$, $K=3$.]{
	%		\resizebox{0.35\linewidth}{!}{%
	%		{\includegraphics[clip, trim=1cm 0.5cm 1cm 1cm, scale=1]{distribution_dist_M_4_N_8_K_3.png}} %gauche bas droite haut
	%	}
	%	}
	\caption{\small Spatial focusing [dB] of the distortion for $K=2$ and $K=3$ users located at red crosses. Axis are in meters units. The distortion becomes more uniformly distributed going from the the single array ($M=1$) to the distributed case ($M=4$) together with an increasing number of users. Moreover, small-scale fading effects can be observed in the distributed case at locations where multiple beams are crossing.}
	\label{fig:spatial_focusing} 
	\vspace{-1em}
\end{figure*}

%	\begin{figure}[t!]
%		\centering 
%		\resizebox{0.9\linewidth}{!}{%
%			{\includegraphics[clip, trim=0cm 0cm 0cm 1cm, scale=1]{CDF_Spatial_Focusing.pdf}} %gauche bas droite haut
%		}
%		\caption{\small Cumulative density function of the spatial focusing of the signal and the distortion. Going from central {\(M=1\)} to distributed massive MIMO {\(M=4\)} and larger number of users make the spatial distribution of distortion more uniform.}
%		\label{fig:spatial_focusing_CDF} 
%		\vspace{-1em}
%	\end{figure}

\subsection{Multi-User Case}\label{subsection:multiple_user}

\subsubsection{Useful Signal Part}\label{sec:multi-user-signal-part}

In the multi-user case, $K$ signals are beamformed in the direction of each \gls{ue} from each array. The useful signal part of (\ref{eq:y_mn_gen}) is given by the term $l=1$. Using (\ref{eq:x_mf}), it is given by
\begin{align*}
b_1 x_{m,n}(t)  = b_1\sum_{k=1}^{K} e^{\jmath \phi_m^{(k)} n+\jmath\psi_m^{(k)}} s^{(k)}(t+\tau_{m}^{(k)}),
\end{align*} 
%and its autocorrelation can be obtained from (\ref{eq:R_x_x_sp}). % as
%\begin{align*}
%|b_1|^2\sum_{k=1}^{K} p^{(k)}e^{\jmath (\phi_m^{(k)}n-\phi_{m'}^{(k)}n') } R(\tau+\tau_{m}^{(k)}-\tau_{m'}^{(k)})\nonumber.
%\end{align*}
%
%Using the same steps as in previous sections, the autocorrelation at the given location can be computed. Finally, taking the Fourier transform, the PSD of the useful signal part at the given location is given by
where we see that each array beamforms in the $K$ directions of the users, \textit{i.e.}, $\phi_m^{(k)}$ for $k=1,...,K$. Using the same steps as in previous sections, the \gls{psd} of the useful signal part at the observer location can be computed as
\begin{align*}
&{S}_{\mathrm{sig}}(\omega)=c_1S(\omega)\sum_{k=1}^K p^{(k)}\\
&\left|\sum_{m=1}^{M}|\beta_m| D_{N_m}(\phi_m^{(k)}- \phi_m)e^{\jmath (\psi_m^{(k)}- \psi_m)+\jmath \omega (\tau_{m}^{(k)}-\tau_{m})}\right|^2.
\end{align*}
The radiation pattern of the signal radiated by one of the arrays is obtained as a particular case of this expression for the array in question, \textit{i.e.}, we set $M=1$ and we drop the array index $m$
\begin{align*}
P_{\mathrm{sig}}(\omega)=S(\omega)c_1|\beta|^2\sum_{k=1}^K p^{(k)}\left|D_{N_m}(\phi^{(k)}- \phi)\right|^2.
\end{align*}
Using the relationship $\phi=\frac{2\pi}{\lambda_c} d\cos(\theta)$, Fig.~\ref{fig:radiation_patterns} plots in blue the array directivity for the useful signal, \textit{i.e.}, normalized with respect to an isotropic radiator. Note that the directivity does not depend on the frequency $\omega$, the PA coefficient $c_1$ and $\beta$. 

%Moreover, one can note that some user interference is possible if beams overlap, as shown in Fig.~\ref{fig:radiation_patterns}(d), where the lower number of antennas induces a large beam width and cannot fully separate the users. This is a well known disadvantage of using matched filtering.

\subsubsection{Distortion Term}

We now focus on the third-order distortion term, \textit{i.e.}, the term of the sum corresponding to $l=3$ in (\ref{eq:y_mn_gen}). Using (\ref{eq:x_mf}), we find that
\begin{align*}
&b_3 x_{m,n}(t)  |x_{m,n}(t)|^{2}= b_3\sum_{k=1}^{K}\sum_{k'=1}^{K}\sum_{k''=1}^{K} e^{\jmath \phi_m^{(k,k',k'')} n}\\
& e^{\jmath \psi^{(k,k',k'')}_m }s^{(k)}(t+\tau_{m}^{(k)})s^{(k')}(t+\tau_{m}^{(k')})(s^{(k'')}(t+\tau_{m}^{(k'')}))^*,
\end{align*}
where $\psi^{(k,k',k'')}_m=\psi_m^{(k)}+\psi_m^{(k')}-\psi_m^{(k'')}$ and $\phi_m^{(k,k',k'')}=\phi_m^{(k)}+\phi_m^{(k')}-\phi_m^{(k'')}$. As compared to the single-user case, the distortion is beamformed not only in the user directions but also several other intermodulation directions~\cite{Hemmi2002}. To clarify this, let us look at the received PSD. Using the same steps as in previous sections, the \gls{psd} of the third-order distortion at the observer location can be computed as
\begin{align*}
{S}_{\mathrm{dis}}^{(1)}(\omega)=&c_3\sum_{k,k',k''}p^{(k)}p^{(k')}p^{(k'')} \sum_{m,m'} |\beta_m||\beta_{m'}| \\
&  e^{\jmath(\psi^{(k,k',k'')}_m-\psi^{(k,k',k'')}_{m'})- \jmath\omega (\tau_m-\tau_{m'})} S^{(3)}_{m,m',k,k'}(\omega)\nonumber\\
&  D_{N_m}(\phi_m^{(k,k',k'')} -\phi_m) (D_{N_{m'}}(\phi_{m'}^{(k,k',k'')} -\phi_{m'}))^*,
\end{align*}
where $ S^{(3)}_{m,m',k,k'}(\omega)=\frac{1}{4\pi^2}e^{\jmath \omega \Delta\tau_{m,m'}^{(k)}} S(\omega)\circledast e^{\jmath \omega \Delta\tau_{m,m'}^{(k')}} S(\omega)\circledast e^{\jmath \omega \Delta\tau_{m,m'}^{(k'')}} S(-\omega)$,
with $\Delta\tau_{m,m'}^{(k)}=\tau_m^{(k)}-\tau_{m'}^{(k)}$.
%We can get a simpler upper bound which neglects small-scale fading/non coherent combining, this is really the worst case when all signals sent to all users from all arrays are combining
%\begin{align*}
%&{S}_{rr,\mathrm{dis},1}(\omega)\leq c_3 S_{ss}^{(3)}(\omega)\sum_{k=1}^{K}\sum_{k'=1}^{K}\sum_{k''=1}^{K}p^{(k)}p^{(k')}p^{(k'')}  \\
%&\left|\sum_{m=1}^{M} \beta_m \frac{\sin\left(\frac{\phi_m-\phi_m^{(k)}}{2}N_m\right)}{\sin\left(\frac{\phi_m^{(k)}+\phi_m^{(k')}-\phi_m^{(k'')} -\phi_m}{2}\right)}  \right|^2\nonumber
%\end{align*}
The distortion radiation pattern of a specific array is obtained as a particular case of this expression for the array in question, \textit{i.e.}, we set $M=1$ and we drop the array index $m$
{\small \begin{align*}
	P_{\mathrm{dis}}^{(1)}(\omega)=& S^{(3)}(\omega)c_3|\beta|^2 \sum_{k,k',k''}p^{(k,k',k'')}| D_{N}(\phi^{(k,k',k'')} -\phi)|^2,
	\end{align*}}
with $p^{(k,k',k'')}=p^{(k)}p^{(k')}p^{(k'')}$. Similarly as for the useful signal, we can define the array directivity of the third-order distortion as $P_{\mathrm{dis}}^{(1)}(\omega,\phi)$ normalized by a isotropically radiated distortion.%\footnote{Note that the distortion directivity does not depend on the frequency $\omega$, the \gls{pa} coefficient $c_3$ and $\beta$.}  
It is shown in dashed red in Fig.~\ref{fig:radiation_patterns}, with related explanations, and in accordance with previous work \cite{Mollen2018TWC}.

The set of beam directions are given by the potential combinations of directions $\phi^{(k,k',k'')}=\phi^{(k)}+\phi^{(k')}-\phi^{(k'')}$, for $k,k',k''=1,...K$. This results in a total of $K^3/2-K^2/2+K$ different directions. As an example, for $K=2$, a total of 4 directions are possible: $\phi^{(1)},\phi^{(2)},2\phi^{(2)}-\phi^{(1)},2\phi^{(1)}-\phi^{(2)}$. %Note that each beam direction should be mapped to a physical angle $\theta$ at each array as $\phi=\frac{2\pi}{\lambda_c} d\cos(\theta) \Longleftrightarrow \theta = \arccos (\phi\lambda_c/(2\pi d))$. 
The power of each beam also depends on the allocated power $p^{(k)}$ to each \gls{ue}. In some cases, directions can combine, resulting in a lower total number of beams. As $N$ decreases, the beam width increases and beam directions are more likely to overlap and to combine, resulting in a more uniform distribution (see Fig.~\ref{fig:radiation_patterns}~(c)~and~(d)). As a comparison with a central massive MIMO system ($M=1$), a distributed massive MIMO system with the same total number of antennas will induce a more uniformly spread distortion since: i) the array gain in main beam directions will be reduced and ii) the beam width will be wider and different beam directions are more likely to overlap. As a result, the distortion becomes uniformly distributed. These effects can be expected from Fig.~\ref{fig:radiation_patterns} by considering a single $32$-antenna array versus four $8$-antenna arrays. Moreover, the distortion gets more uniformly beamformed with increasing number of distinct \glspl{ue} directions.

\vspace{-0.65em}

\section{Spatial Distribution in a Cell}\label{section:Case_vstudy}

In this section we study the spatial distribution of the signal and the third-order distortion, for a $100m\times 100m$ square cell, as shown in Fig.~\ref{fig:spatial_focusing}. We compare the performance of a $M=4 \times N=8$ distributed system with a $M=1 \times N=32$ central system. To study the radiated power at observer locations, we sample the cell with a spatial step of $\lambda_c/2$, which captures small scale fading effects. A path loss exponent of \num{2.5} is used. The figure of merit under study is the spatial focusing, defined as the power radiated at a certain location versus an ideally uniformly distributed power radiation. In other words, ${S}_{\mathrm{sig}}(\omega)$ and ${S}_{\mathrm{dis}}^{(1)}(\omega)$ are respectively normalized by their average computed on the whole \SI{100}{\meter}$\times$\SI{100}{\meter} cell area. The spatial focusing does not depend on the \gls{pa} parameters ($c_1$ and $c_3$) and can be seen as a spatial generalization of the array directivity. The spatial focusing is evaluated at carrier frequency $f_c=1$ GHz, for a raised cosine \glspl{psd} $S(\omega)$. Fig.~\ref{fig:spatial_focusing} clearly shows that the distortion becomes more uniformly distributed going from: i) the single array $M=1$ to the distributed case ($M=4$) and ii) $K=2$ to $K=3$ users. Moreover, one can notice in the distributed case the so-called small scale fading effect, previously described and resulting from the non-coherent combining of the transmissions from the $M$ arrays. %Fig.~\ref{fig:spatial_focusing_CDF} plots the cumulative density function of the signal and distortion for the same scenarios. It further confirms that going from central to distributed massive MIMO and larger number of users make the spatial distribution of power more uniform.

%\liesbet{I would not go very ambitious here, maybe just formulate a few use cases, to be discussed}

%\gilles{ultimate cell-free case}
%\gilles{single user/multi-user/NloS cases}
%\gilles{cdf of oob?}
 	\vspace{-0.5em}

\section{Conclusion}\label{section:conclusion}

We have shown in this paper that the distortion due to nonlinear \gls{pa} is not always uniformly distributed in space. In the single-user \gls{los} case, it coherently adds up at the user location. In the few users case, with a few distinct beam directions, the signals will add up at the user locations plus several others locations. As the number of beam directions increases, it quickly becomes close to uniformly distributed. As a comparison with a massive MIMO system having the same total number of antennas, the distortion in distributed massive MIMO is considerably more uniformly distributed in space. Moreover, the potential coherent combining is contained in a beamspot rather than in generic directions and it is subject to small-scale fading effects. As a general conclusion, we can expect that going distributed allows working significantly closer to saturation, directly improving the \gls{pa} efficiency. Future works will target a quantitative assessment of this improvement.% and the extension of the approach to general multipath channels, different types of precoders and non uniform power allocation among the arrays. Another aspect is the extension towards a realistic non-omnidirectional, potentially suffering from coupling and polarization mismatch, 3D array pattern. A final aspect is the particularization of the signal model to a CP-OFDM system, taking into account a per-subcarrier, per-OFDM symbol precoding.

% \francois{we can discuss future extensions such as power scaling Laws (how to change the PA when number of antennas varies), back off strategy, power allocation, generalize single-carrier formulation to OFDM, MF beamformer, omnidirectional antenna without coupling and no cross-polarization, only 2D pattern, LOS channel, single-user or few users, no power allocation between arrays} \gilles{imperfect CSI?}
% \liesbet{add a 'what it means' in practise if we can work xx dB closer to saturation results in yy improved energy in the PA stage and will also reduce PA cost}
% \gilles{Class A/B amplifiers have a theoretical efficiency~\cite{cripps2006rf} of \(\eta = \frac{\pi}{4} \sqrt{b},\) with \(b\) a given back-off. \(b\) is defined between \(0\) and \(1\). At a backoff of \(1\), the \gls{pa} works at maximum efficiency (\(\eta \approx 78\text{\%}\)). Reformulating: \(\eta = \frac{\pi}{4} 10^{(-\text{bo\_db}/20)}\). Hence, the power efficiency is proportional to the square root of the back-off power. } 

% \begin{figure}
%   \begin{tikzpicture}
%         \begin{axis}[
%         %axis lines = left,
%         xlabel = back-off (dB),
%         ylabel = {$\eta$ (\%)},
%         ]
%             \addplot[blue, ultra thick, domain=0:20]{100*(pi/4)*(10^(-x/20))};
%             \addlegendentry{$\eta$ (\%)}
%             \addplot[red, ultra thick, domain=0:20]{100*(pi/4)*(1-(10^(-x/20)))};
%             \addlegendentry{$\eta_{\text{loss}}$ (\%) wrt 0dB}
%         \end{axis}
%     \end{tikzpicture}
% \end{figure}

%\section{Appendix}

\vspace{-0.5em}

\section*{Acknowledgment}
{\ The research reported herein was partly funded by Huawei and the F.R.S.-FNRS.}

\vspace{-0.5em}
%\footnotesize
\scriptsize 
\bibliographystyle{IEEEtran}
\bibliography{IEEEabrv,IEEEreferences}

\end{document}